\documentclass[a4paper,fleqn,usenatbib]{mnras}
\usepackage{subfigure}
\usepackage{rotating}
\usepackage{newtxtext,newtxmath}
\usepackage[T1]{fontenc}
\usepackage{ae,aecompl}
\usepackage{tikz}
\usepackage{nccmath}
\usepackage{multicol, blindtext}

\usepackage{graphicx}	
\usepackage{amsmath}	
\usepackage{float}		
\usepackage{placeins}
\usepackage{array}
\usepackage{epsfig}

\newcommand{\Ms}{M$_{\odot}$}

\title[Early QSO Radio Emission]{Radio Emission from the First Quasars at $z =$ 6 - 15}
\author[Latif et al.]{Muhammad A. Latif,$^{1}$\thanks{E-mail: latifne@gmail.com} 
Daniel J. Whalen,$^2$
Mar Mezcua$^{3,4}$
\\
\\
$^1$Physics Department, College of Science, United Arab Emirates University (UAEU), PO Box 15551, Al-Ain, UAE \\
$^2$Institute of Cosmology and Gravitation, University of Portsmouth, Dennis Sciama Building, Portsmouth PO1 3FX, UK \\
$^3$Institute of Space Sciences (ICE, CSIC), Campus UAB, Carrer de Magrans, 08193 Barcelona, Spain \\
$^4$Institut d'Estudis Espacials de Catalunya (IEEC), Carrer Gran Capit\`{a}, 08034 Barcelona, Spain \\
}

\date{Accepted XXX. Received YYY; in original form ZZZ}

\begin{document}
\pagerange{\pageref{firstpage}--\pageref{lastpage}}
\maketitle

\begin{abstract}

Nearly 300 quasars have now been found at $z >$ 6, including nine at $z >$ 7.  They are thought to form from the collapse of supermassive primordial stars to 10$^4$ - 10$^5$ \Ms\ black holes at $z \sim$ 20 - 25, which then rapidly grow in the low-shear environments of rare, massive halos fed by strong accretion flows.  Sensitive new radio telescopes such as the Next-Generation Very Large Array (ngVLA) and the Square Kilometer Array (SKA) could probe the evolution of these objects at much earlier times.  Here, we estimate radio flux from the first quasars at $z \sim$ 6 - 15 at 0.5 - 12.5 GHz.  We find that SKA and ngVLA could detect a quasar like ULAS J1120+0641, a 1.35 $\times$ 10$^9$ \Ms\ black hole at $z =$ 7.1, at much earlier stages of evolution, $z \sim$ 14 - 16, with 100 hr integration times in targeted searches.  The advent of these new observatories, together with the \textit{James Webb Space Telescope} (\textit{JWST}), \textit{Euclid}, and the \textit{Roman Space Telescope} (\textit{RST}), will inaugurate the era of $z \lesssim$ 15 quasar astronomy in the coming decade.
    
\end{abstract}


\section{Introduction}

Nearly 300 quasars have now been discovered at $z >$ 6, including nine at $z > 7$ (e.g., \citealt {yang20,wang21} -- see \citealt{fan23} for a recent review).  The seeds of these quasars are thought to be direct-collapse black holes (DCBHs) from the collapse of 10$^4$ - 10$^5$ \Ms\ primordial stars \citep{wf12,hos13,tyr21a,herr23a}.  But to reach 10$^9$ \Ms\ by $z \sim$ 6 - 7 they must also be located in rare, massive haloes at the nexus of strong accretion flows in low-shear environments \citep{ten18,lup21,vgf21,latif22b}.  A number of studies have shown that DCBHs could be found at $z \gtrsim$ 20 in the near infrared (NIR) by \textit{JWST} \citep{pac15,wet20b} and $z \lesssim$ 15 by \textit{Euclid}, and the \textit{RST} \citep{vik22a}.  

\begin{figure*}
\begin{center}
\begin{tabular}{cc}
\includegraphics[width=0.42\linewidth]{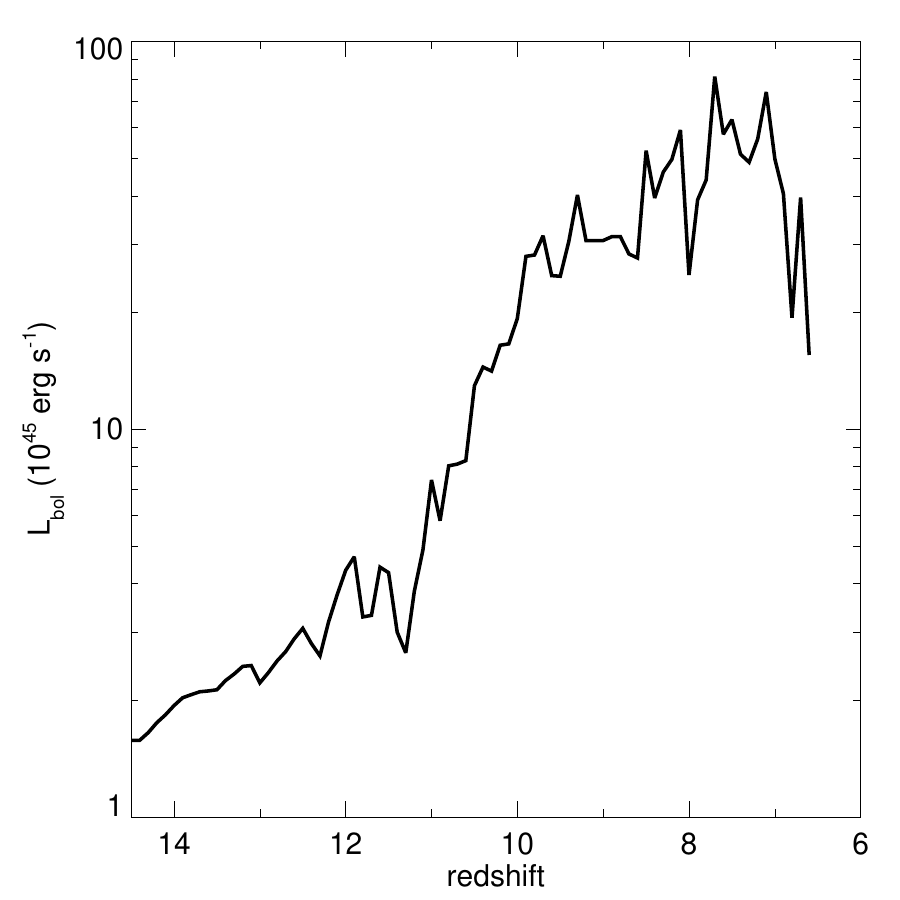}
\includegraphics[width=0.42\linewidth]{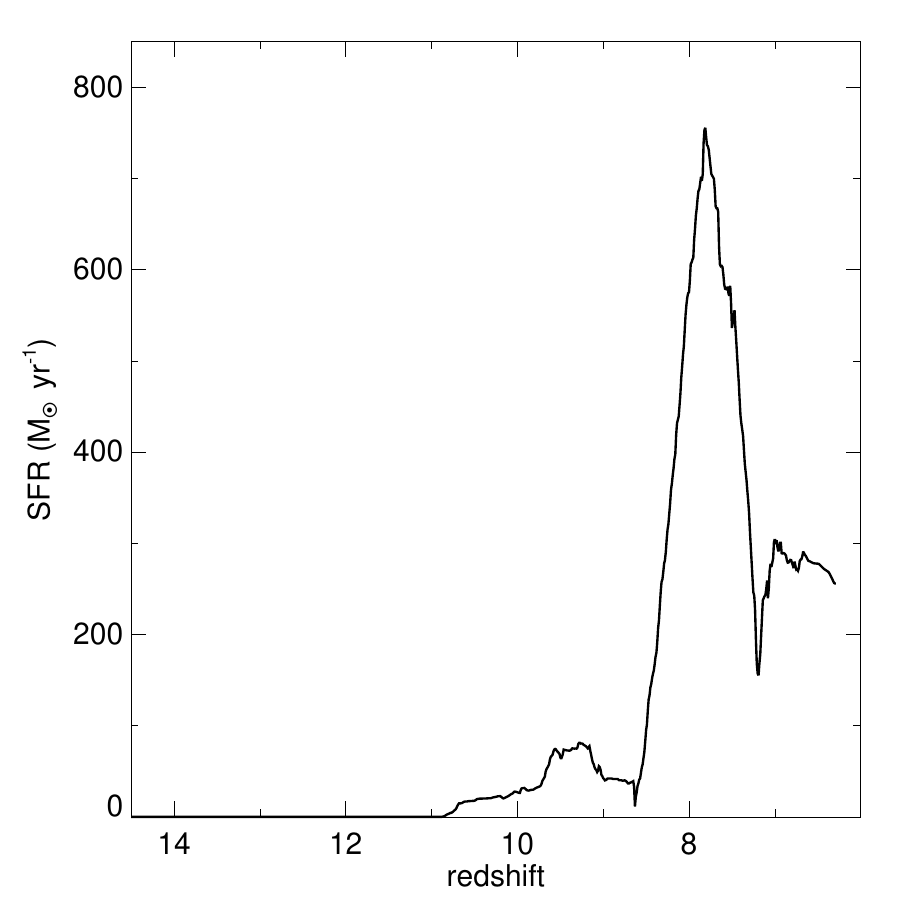} \\
\end{tabular}
\end{center}
\caption{Bolometric luminosities, $L_{\mathrm{bol}}$, for the quasar (left) and star formation rates in its host galaxy (right) as functions of redshift from \citet{smidt18}.}
\label{fig:lbol}
\end{figure*}

Radio-loud quasars (i.e., with radio fluxes 10 times greater than their optical blue-band emission -- \citealt{glou22,ban23}) have now also been found at $z >$ 6, such as J0309+2717 \citep[$z = $ 6.10;][]{bd20}, J1427+3312 \citep[$z = $ 6.12;][]{mcg06}, J1429+5447 \citep[$z = $ 6.18;][]{wil10}, P172+18 \citep[$z = $ 6.82;][]{ban21}, and VIK J2318+3113 \citep[$z = $ 6.444;][]{zng22}.  They typically have BH masses $\gtrsim$ 10$^8$ \Ms\ and grow at nearly the Eddington rate \citep[e.g.,][]{ban21}. Except for J0309+2717 \citep[classified as a blazar;][]{spng20}, the radio emission from these sources is dominated by compact cores and/or small-scale jets ($<$ 1 kpc).  Although DCBH detections by SKA and ngVLA will be limited to $z \lesssim$ 8, \citep{wet20a,wet21a}, these observatories in principle could detect more massive quasars at well above this redshift, at much earlier stages of their evolution.  Here, we estimate radio flux from a $z \sim$ 7 quasar like ULAS J1120+0641 \citep{mort11} at earlier stages of growth, up to $z \sim$ 15. Our calculations are timely, given the recent discovery of a lensed 4 $\times$ 10$^7$ \Ms\ BH candidate in UHZ1 at $z =$ 10.3 \citep{akos23}.  In Section 2 we discuss how we calculate radio emission from the BH and star-forming regions in its host galaxy.  We present radio fluxes for a number of SKA and ngVLA bands as a function of redshift in Section 3 and then conclude in Section 4.

\section{Numerical Method}

We estimate quasar radio fluxes from fundamental planes (FPs) of BH accretion with bolometric luminosities taken from \citet{smidt18}, who were the first to reproduce the properties of a $z \sim$ 7 quasar in a cosmological simulation (left panel of Figure~\ref{fig:lbol}).  We use second-year \textit{Planck} cosmological parameters:  $\Omega_{\mathrm M} = 0.308$, $\Omega_\Lambda = 0.691$, $\Omega_{\mathrm b}h^2 = 0.0223$, $\sigma_8 =$ 0.816, $h = $ 0.677 and $n =$ 0.968 \citep{planck2}.

\subsection{BH Radio Flux}

FPs of BH accretion are empirical relationships between the mass of a BH, $M_\mathrm{BH}$, its nuclear X-ray luminosity at 2 - 10 keV, $L_\mathrm{X}$, and its nuclear radio luminosity at 5 GHz, $L_\mathrm{R}$ (\citealt{merl03}; see \citealt{mez18} for a brief review).  They cover six orders of magnitude in BH mass, including down to intermediate-mass black holes \citep[$< 10^5$ \Ms;][]{gul14}  To estimate the flux from the BH in a radio band in the Earth frame we first use the FP to calculate $L_\mathrm{R}$ in the source frame, which depends on $M_\mathrm{BH}$ and $L_\mathrm{X}$.  We obtain $L_\mathrm{X}$ from $L_{\mathrm{bol}}$ with Equation 21 of \citet{marc04},
\begin{equation}
\mathrm{log}\left(\frac{L_\mathrm{bol}}{L_\mathrm{X}}\right) = 1.54 + 0.24 \mathcal{L} + 0.012 \mathcal{L}^2 - 0.0015 \mathcal{L}^3,
\end{equation}
where $L_\mathrm{bol}$ is in units of solar luminosity and $\mathcal{L} = \mathrm{log} \, L_\mathrm{bol} - 12$.  $L_\mathrm{R}$ can then be calculated from $L_\mathrm{X}$ with an FP of the form
\begin{equation}
\mathrm{log} \, L_\mathrm{R} (\mathrm{erg \, s^{-1}})= \alpha \, \mathrm{log} \, L_\mathrm{X} (\mathrm{erg \, s^{-1}}) + \beta \, \mathrm{log} \, M_\mathrm{BH} (\mathrm{M}_{\odot}) + \gamma,
\end{equation}
where $\alpha$, $\beta$ and $\gamma$ for FPs from \citet[][MER03]{merl03}, \citet[][KOR06]{kord06}, \citet[][GUL09]{gul09}, \citet[][PLT12]{plot12}, and \citet[][BON13]{bonchi13} are listed in Table~1 of \citet{wet21a}.  We also include the FP of Equation 19 in \citet[][GUL19]{gul19}, 
\begin{equation}
R \, = \, -0.62 + 0.70 \, X + 0.74 \, \mu,
\end{equation}
where $R =$ log($L_\mathrm{R}/10^{38} \mathrm{erg/s}$), $X =$ log($L_\mathrm{X}/10^{40} \mathrm{erg/s}$) and $\mu =$ log($M_\mathrm{BH}/10^{8}$\Ms). 

\begin{figure*} 
\begin{center}
\begin{tabular}{cc}
\includegraphics[width=0.42\linewidth]{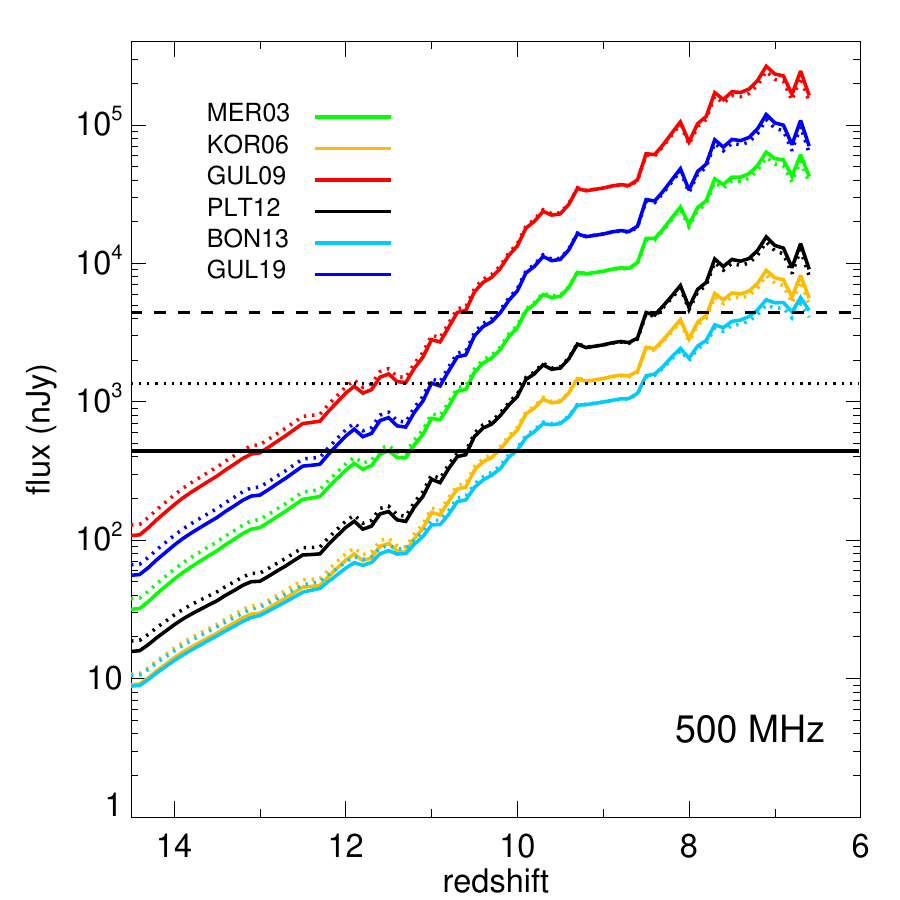}  &
\includegraphics[width=0.42\linewidth]{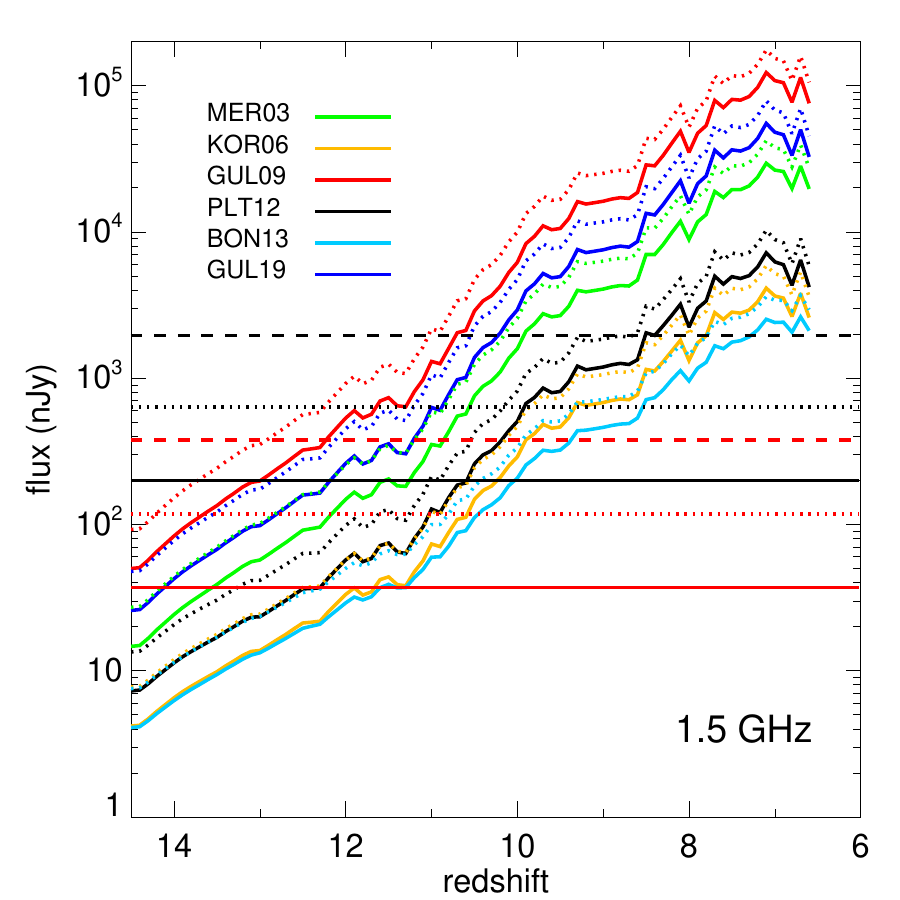}  \\
\includegraphics[width=0.42\linewidth]{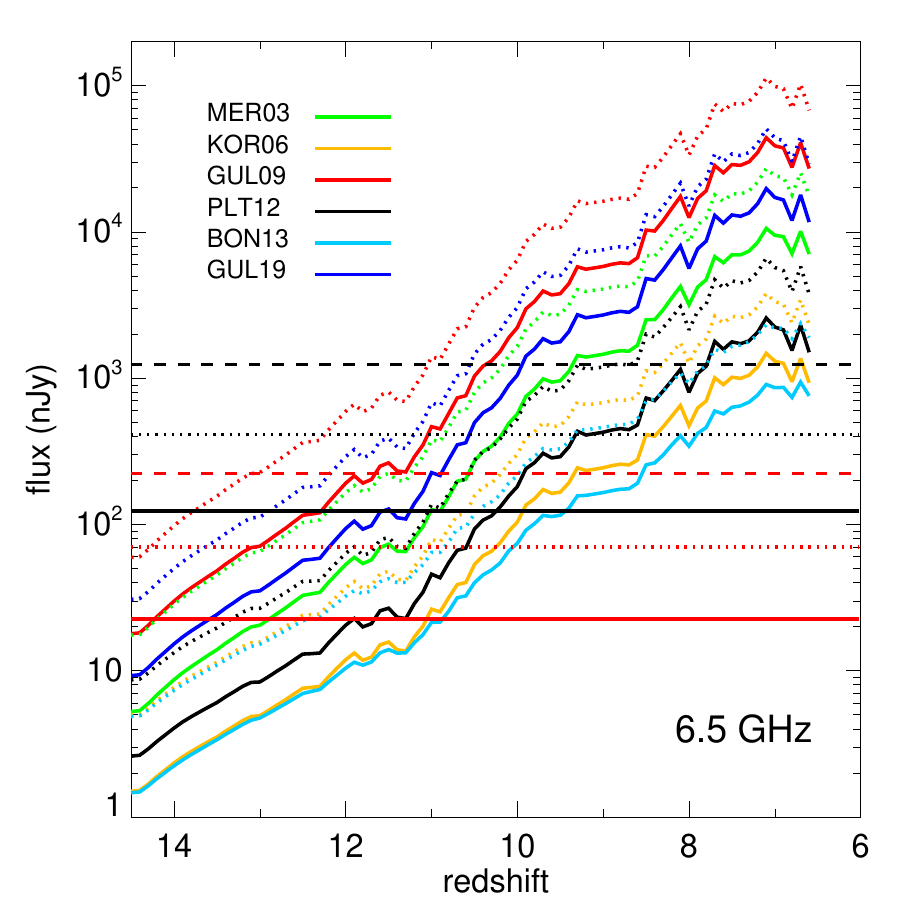}  &
\includegraphics[width=0.42\linewidth]{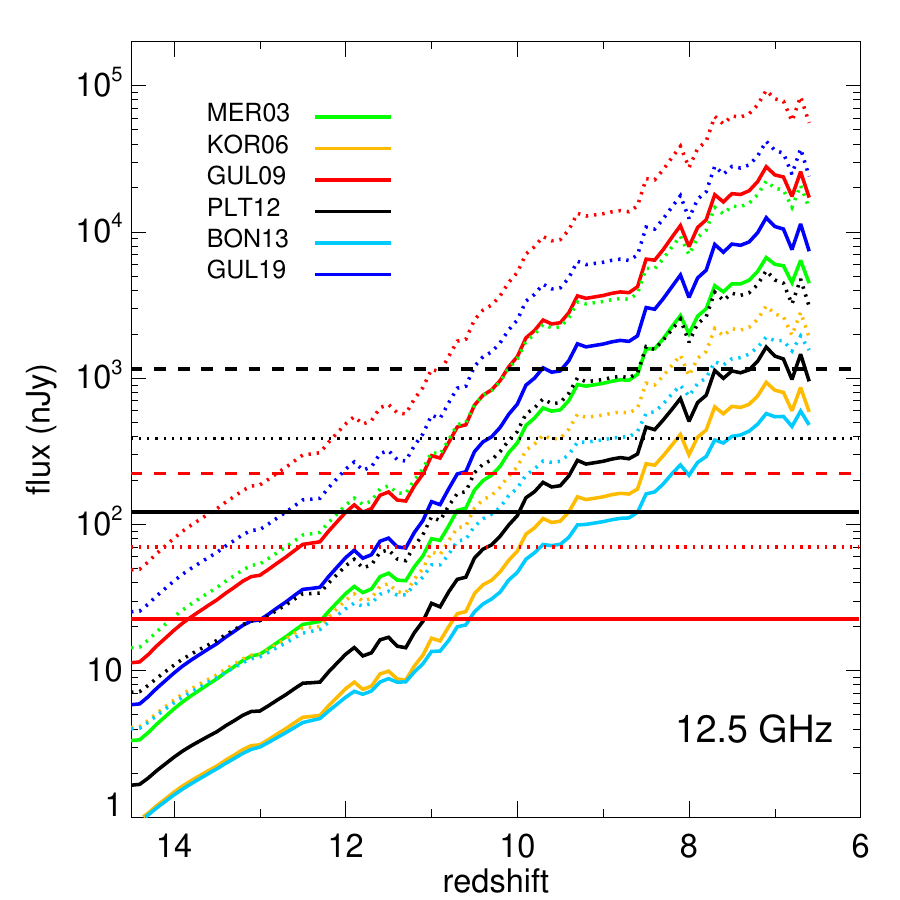}  
\end{tabular}
\end{center}
\caption{BH radio fluxes for $\alpha =$ 0.7 (solid) and $\alpha =$ 0.3 (dotted) from $z =$ 6 - 14.5 at 500 MHz, 1.5 GHz, 2.5 GHz and 6.5 GHz. The dashed, dotted and solid horizontal lines are the detection limits for 1 hr, 10 hr and 100 hr integration times for SKA (black) and ngVLA (red) listed in Table~\ref{tab:sens}.}
\vspace{0.2in}
\label{fig:flux} 
\end{figure*}

Since radio flux from a supermassive black hole (SMBH) that is cosmologically redshifted into a given observer band today does not originate from 5 GHz in the source frame at high redshifts, we calculate it from $L_\mathrm{R} =$ $\nu L_{\nu}$, assuming that the spectral luminosity $L_{\nu} \propto \nu^{-\alpha}$.  Here, we consider $\alpha =$ 0.7 \citep{ccb02} and $\alpha =$ 0.3 \citep{glou21} for a reasonble range of spectral types.  The spectral flux at $\nu$ in the observer frame is then calculated from the spectral luminosity at $\nu'$ in the rest frame from
\begin{equation}
F_\nu = \frac{L_{\nu'}(1 + z)}{4 \pi {d_\mathrm L}^2},
\end{equation}
where $\nu' = (1+z) \nu$ and $d_\mathrm L$ is the luminosity distance.    

\subsection{H II Region Radio Emission}

Thermal bremsstrahlung emission in H II regions can produce continuum flux whose spectral radio density can be derived from the ionizing photon emission rate in the H II region, $Q_{\mathrm{Lyc}}$, from
\begin{equation}
L_{\nu} \, \lesssim \, \left(\frac{Q_{\mathrm{Lyc}}}{6.3 \times 10^{52} \, \mathrm{s^{-1}}}\right) \left(\frac{T_{\mathrm{e}}}{10^4 \mathrm{K}}\right)^{0.45} \left(\frac{\nu}{\mathrm{GHz}}\right)^{-0.1}
\end{equation}
in units of 10$^{20}$ W Hz$^{-1}$ \citep{con92}, where $Q_{\mathrm{Lyc}} =$ SFR (\Ms yr$^{-1}$) $/$ $1.0 \times 10^{-53}$ \citep{ken98} and SFR is the star formation rate.  We estimate the radio continuum from star-forming regions in the host galaxy of the SMBH with the SFRs from \citet{smidt18} shown in the right panel of Figure~\ref{fig:lbol}, taking $T_{\mathrm{e}} =$ 10$^4$ K. 

\subsection{Supernova Radio Flux}

Radio flux from early quasars could include synchrotron emission from young supernova (SN) remnants in the host galaxy.  However, since SFRs in the galaxy are quite low until $z \sim$ 10 - 11, we would not expect SNe to contribute much to the flux prior to these redshifts.  \citet{con92} estimate the non-thermal radio luminosity from SNe in normal star-forming galaxies today to be
\begin{equation}
\left(\frac{L_{\mathrm{N}}}{\mathrm{W \, Hz^{-1}}}\right) \, \sim \, 4.4 \times 10^{34} \left(\frac{\nu}{\mathrm{GHz}}\right)^{-\beta} \left[\frac{\mathrm{SFR}(M > 5 \, \mathrm{M_{\odot}})}{\mathrm{M_{\odot}} \, \mathrm{yr^{-1}}} \right],
\label{eq:SNe}
\end{equation}
where $\beta \sim$ 0.8 is the spectral index.  But Equation~\ref{eq:SNe} would not be valid in the halo at early times because Pop III core-collapse SNe \citep{jet09b,latif22a} only produced nJy radio fluxes \citep{mw12} in the diffuse H II regions of high redshifts \citep{wan04,wet08a}, well below the detection limit of any planned survey.  However, they may later approach those of Equation~\ref{eq:SNe} at lower redshifts and near-solar metallicities so we adopt it for simplicitiy.  Since the IMF of the stars in the host galaxy is unknown, we simply assume that they are above 5 \Ms\ at all redshifts, so this constitutes a severe upper limit to the true SN flux.

\section{Results}

We show $F_\nu$ at 500 MHz, 1.5 GHz, 6.5 GHz and 12.5 GHz at $z =$ 6 - 14.5 for all six FPs and both choices of spectral index in Figure~\ref{fig:flux} along with the sensitivity limits for 1 hr, 10 hr, and 100 hr pointings by SKA and ngVLA listed in Table~\ref{tab:sens}.  For comparison, the most sensitive surveys by the Low-Frequency Array (LOFAR) will reach a few $\mu$Jy below 1 GHz.  The BH fluxes are higher at lower frequencies, where the estimates range from $\sim$ 9 - 120 nJy at $z =$ 14.5 to $\sim$ 5 $\mu$Jy - 0.2 mJy at $z =$ 6 at 500 MHz.  At 6.5 GHz the fluxes are lower, $\sim$ 1.5 - 60 nJy at $z =$ 14.5 to $\sim$ 0.8 $\mu$Jy - 70 $\mu$Jy at $z =$ 6.  The choice of spectral index does not have much impact on the flux at 500 MHz but can enhance it by factors of up to 5 at higher wavelengths, where $\alpha =$ 0.3 produces higher fluxes at all redshifts.

\begin{table}
\caption{
Planned sensitivities (nJy/beam) for SKA1 at 500 MHz, 1.5 GHz, 6.5 GHz, and 12.5 GHz for 1 hr, 10 hr, and 100 hr integration times \citep[top three rows; see Table 3 of][]{ska} and ngVLA  \citep[bottom three rows;][]{pr18}.
}
\label{tab:sens}
\begin{center}
\begin{tabular}{lcccc}
\hline
\\
 & 500 MHz & 1.5 GHz & 6.5 GHz  & 12.5 GHz \\
\hline
\\
1 hr     & 4400  & 2000 & 1300 & 1200  \\
10 hr   & 1391  & 632   & 411   &  379   \\
100 hr & 440    & 200   & 130   & 120    \\
\\
\hline
\\
1 hr     & ---  & 382 & 220 & 220  \\
10 hr   & ---  & 121 & 70   &  70   \\
100 hr & ---  & 38   & 22   &  22   \\ 
\\
\hline 
\vspace{-0.3in}
\end{tabular}
\end{center}
\end{table}

For $\alpha =$ 0.7 and 0.3, the most optimistic FPs predict that 1 hr, 10 hr and 100 hr pointings by SKA could detect the quasar at 500 MHz at $z \lesssim$ 10.5, 12, and 13, respectively.  At 1.5 GHz, the same SKA times could detect it at $z \lesssim$ 11, 12, and 13 - 13.5 (hereafter, upper and lower limits in $z$ are for $\alpha =$ 0.3 and 0.7, respectively). The ngVLA could detect the quasar at $z \lesssim$ 12 - 13, 13.5 - 14.2 , and 15 - 15.5 if one extrapolates the plots to redshifts higher than those for which we had bolometric luminosities.  At 6.5 GHz, SKA could detect the BH at $z \lesssim$ 10 - 11, 11 - 12, and 12 - 14 with 1 hr, 10 hr and 100 hr integration times and ngVLA could detect it at 12 - 13, 13.5 - 14.2, and 14.2 - 15.5 if one again extrapolates the plots to higher $z$.  At 12.5 GHz, SKA could detect the quasar at $z \lesssim$ 10 - 11, 11 - 12, and 12 - 13.5 with 1 hr, 10 hr and 100 hr pointings and ngVLA could detect it at 11 - 12.8, 12.5 - 14, and 14 - 15.2.  LOFAR could observe the quasar at $z \lesssim$ 10 at 500 MHz.

\begin{figure}
\centering
\includegraphics[width=0.45\textwidth]{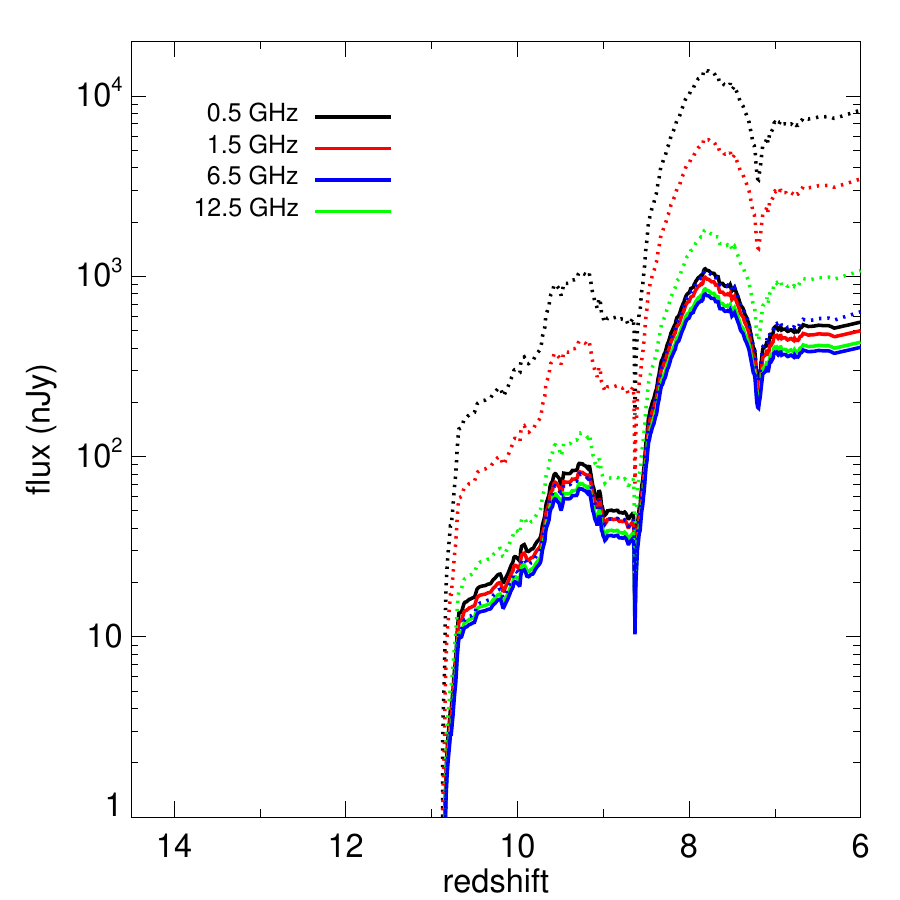}
\caption{Left: radio emission at 0.5 - 12.5 GHz due to thermal bremsstrahlung from H II regions in star-forming regions in the host galaxy of the BH (solid lines).  Right: non-thermal synchrotron emission from SNe (dotted lines).}
\label{fig:HII}
\end{figure}

We show H II region radio fluxes for the host galaxy in Figure~\ref{fig:HII}.  They are similar in strength because of their weak dependence on frequency.  There is essentially no radio emission from the host galaxy until the onset of star formation at $z \sim$ 11. It then rises from $\sim$ 1 nJy at $z =$ 10.8 to a first peak of $\sim$ 10$^2$ Jy at $z =$ 9 - 10 and then a second peak of $\sim$ 10$3$ Jy at $z =$ 8. The peaks and dips in the radio flux in all four bands correspond to those in the SFRs in Figure~\ref{fig:lbol}.  In reality, there is almost certainly some radio emission from the host galaxy at $z >$ 11 because \citet{smidt18} could not resolve resolve the formation of smaller star-forming regions at earlier times.  However, as discussed in the previous section, this emission is expected to be at most a few nJy in the primordial gas.  Radio emission due to thermal bremsstrahlung depends on electron temperatures and densities, so our fluxes should be taken as upper limits because H II regions in local galaxies, from which our estimates are taken, probably have higher densities than those in the host galaxy.  The H II region flux at its two peaks is at most 10\% of the flux from the SMBH.

The SN fluxes are 2 - 10 times higher than the H II region fluxes at the same frequencies at a given redshift.  They exhibit the same dependence on SFR as the H II region emission.  They somewhat exceed the two most pessimistic BH fluxes but the more optimistic BH fluxes can be 10 - 30 times higher in all four frequency bands.  However, as discussed earlier, because we assume that all star formation results in SNe, these should be taken to be extreme upper limits.  The BH flux is likely to be much higher than the SN and H II region fluxes combined at $z \lesssim$ 11.  They could could make the quasar easier to detect when present but not increase the redshifts at which it could be found.  

\section{Discussion and Conclusion}

The first quasars are better candidates for radio followup after being detected at other wavelengths than being discovered in blind radio surveys because their areal densities on the sky are so small, $\sim$10$^{-3}$ deg$^{-2}$ at $z >$ 7 \citep{wang21,nan22}.  The SKA-MID and ultra deep surveys will reach sensitivities of 200 nJy and 50 nJy \citep[Table 1 of][]{ska1} so they could in principle detect quasars at $z \lesssim$ 14, but it is unlikely that any would fall in the 10 - 30 deg$^2$ and 1 deg$^2$ patches of the sky they will cover.  NgVLA radio surveys will reach sensitivities of 45 nJy at 3.5 - 12.3 GHz and 78 nJy at 1.2 - 3.5 GHz with 24 hr integration times, but with typical allocations of a few thousand hours a survey at this depth would cover $\lesssim$ 100 deg$^2$ and probably not find BHs at high redshifts.

The properties of our quasar are similar to those of other $z >$ 7 BHs \citep[see Table 1 and Figure 8 of][]{fan23}, so our fluxes are likely typical of SMBHs of this era.  Many quasars at $z >$ 6 have larger masses so our estimates could be on the low side at these redshifts.  Radio measurements by \citet{mom14} impose 3$\sigma$ upper limits of 23 $\mu$Jy on the 1.4 GHz rest-frame flux of J1120, which is consistent with the lower end of the FPs such as BON13 and our H II region continuum fluxes, which are about 20 $\mu$Jy.  The higher end of our radio fluxes are on par with those recently reported for several $z \sim$ 6 quasars.  \citet{ban21} measured flux densities of 0.31 - 2.9 mJy at 1.4 GHz for J2228+0110, J1427+3312, J1429+5447 and P172+18 at $z =$ 5.95 - 6.82.  \citet{zng22} report a flux density of 5.5 mJy for the radio-loud AGN VIK J2318-3113 at $z =$ 6.44.  These fluxes are a factor of a few higher than those of the more optimistic FPs we considered but these sources are thought to be blazars that are unusually radio-loud.  In particular, very-long baseline interferometry determined that the radio emission from J2318-3113 originated from a nuclear region about 1 pc across, indicating that the emission was due to the BH itself and not radio jets.  

If the quasar has jets there could be additional radio flux due to synchrotron emission from its lobes that we do not consider here.  However, except for one or two rare cases in which compact jets have been observed from high-redshift quasars on scales of a kpc or less \citep{mom18,con21}, jets are not expected for most of these sources.  They have generally been observed at $L_{\mathrm{bol}} \lesssim 0.01 \, L_{\mathrm{Edd}}$ and $L_{\mathrm{bol}} \gtrsim L_{\mathrm{Edd}}$ and SMBHs must grow at rates of 0.1 - 1 ${\dot{M}}_{\mathrm{Edd}}$ to become quasars by $z \sim$ 6.  The BH in our model accretes at 0.2 $L_{\mathrm{Edd}}$ - 0.9 $L_{\mathrm{Edd}}$ over its lifetime.  If a jet did form, the cosmic microwave background (CMB) would probably quench its emission \citep{gh14,fg14} but not the FP fluxes because they originate from the central region of the quasar.

SKA and ngVLA will probe the properties of the first quasars at much earlier stages of evolution in the coming decade, discriminating between seeding mechanisms such as birth as DCBHs or prompt formation during galaxy mergers \citep[e.g.,][]{may10}.  The ngVLA will be the better observatory for following up on quasar candidates previously discovered with other telescopes because it can detect them at up to $z \sim$ 16 with 100 hr pointing times while  SKA will be limited to $z \sim$ 14.  They will also complement NIR surveys by eliminating BH impostors, since strong radio emission would not be expected from active SF in early galaxies alone.  SKA and the ngVLA will, together with \textit{JWST}, \textit{Euclid} and the \textit{RST}, will inaugurate the era of $z \sim$ 15 quasar astronomy in the coming decade.

\section*{Acknowledgements}

We thank the anonymous referee, whose critique improved the quality of this work.  M. A. L. thanks the UAEU for funding via UPAR grant No. 31S390 and startup grant No 31S372.  This work was also supported by the program Unidad de Excelencia Mar\'ia de Maeztu CEX2020-001058-M.

\section{Data Availability Statement}

The data in this study will be made available upon request to the corresponding author.

\bibliographystyle{mnras}
\bibliography{refs} 

\begin{thebibliography}{}
\makeatletter
\relax
\def\mn@urlcharsother{\let\do\@makeother \do\$\do\&\do\#\do\^\do\_\do\%\do\~}
\def\mn@doi{\begingroup\mn@urlcharsother \@ifnextchar [ {\mn@doi@}
  {\mn@doi@[]}}
\def\mn@doi@[#1]#2{\def\@tempa{#1}\ifx\@tempa\@empty \href
  {http://dx.doi.org/#2} {doi:#2}\else \href {http://dx.doi.org/#2} {#1}\fi
  \endgroup}
\def\mn@eprint#1#2{\mn@eprint@#1:#2::\@nil}
\def\mn@eprint@arXiv#1{\href {http://arxiv.org/abs/#1} {{\tt arXiv:#1}}}
\def\mn@eprint@dblp#1{\href {http://dblp.uni-trier.de/rec/bibtex/#1.xml}
  {dblp:#1}}
\def\mn@eprint@#1:#2:#3:#4\@nil{\def\@tempa {#1}\def\@tempb {#2}\def\@tempc
  {#3}\ifx \@tempc \@empty \let \@tempc \@tempb \let \@tempb \@tempa \fi \ifx
  \@tempb \@empty \def\@tempb {arXiv}\fi \@ifundefined
  {mn@eprint@\@tempb}{\@tempb:\@tempc}{\expandafter \expandafter \csname
  mn@eprint@\@tempb\endcsname \expandafter{\@tempc}}}

\bibitem[\protect\citeauthoryear{{Ba{\~n}ados} et~al.,}{{Ba{\~n}ados}
  et~al.}{2021}]{ban21}
{Ba{\~n}ados} E.,  et~al., 2021, \mn@doi [\apj] {10.3847/1538-4357/abe239},
  \href {https://ui.adsabs.harvard.edu/abs/2021ApJ...909...80B} {909, 80}

\bibitem[\protect\citeauthoryear{{Ba{\~n}ados} et~al.,}{{Ba{\~n}ados}
  et~al.}{2023}]{ban23}
{Ba{\~n}ados} E.,  et~al., 2023, \mn@doi [\apjs] {10.3847/1538-4365/acb3c7},
  \href {https://ui.adsabs.harvard.edu/abs/2023ApJS..265...29B} {265, 29}

\bibitem[\protect\citeauthoryear{{Belladitta} et~al.,}{{Belladitta}
  et~al.}{2020}]{bd20}
{Belladitta} S.,  et~al., 2020, \mn@doi [\aap] {10.1051/0004-6361/201937395},
  \href {https://ui.adsabs.harvard.edu/abs/2020A&A...635L...7B} {635, L7}

\bibitem[\protect\citeauthoryear{{Bogdan} et~al.,}{{Bogdan}
  et~al.}{2023}]{akos23}
{Bogdan} A.,  et~al., 2023, \mn@doi [] {10.48550/arXiv.2305.15458}, \href
  {https://ui.adsabs.harvard.edu/abs/2023arXiv230515458B} {p. arXiv:2305.15458}

\bibitem[\protect\citeauthoryear{{Bonchi}, {La Franca}, {Melini}, {Bongiorno}
  \& {Fiore}}{{Bonchi} et~al.}{2013}]{bonchi13}
{Bonchi} A.,  {La Franca} F.,  {Melini} G.,  {Bongiorno} A.,   {Fiore} F.,
  2013, \mn@doi [\mnras] {10.1093/mnras/sts456}, \href
  {http://adsabs.harvard.edu/abs/2013MNRAS.429.1970B} {429, 1970}

\bibitem[\protect\citeauthoryear{{Braun}, {Bonaldi}, {Bourke}, {Keane}  \&
  {Wagg}}{{Braun} et~al.}{2019}]{ska}
{Braun} R.,  {Bonaldi} A.,  {Bourke} T.,  {Keane} E.,   {Wagg} J.,  2019,
  \mn@doi [] {10.48550/arXiv.1912.12699}, \href
  {https://ui.adsabs.harvard.edu/abs/2019arXiv191212699B} {p. arXiv:1912.12699}

\bibitem[\protect\citeauthoryear{{Condon}}{{Condon}}{1992}]{con92}
{Condon} J.~J.,  1992, \mn@doi [\araa] {10.1146/annurev.aa.30.090192.003043},
  \href {https://ui.adsabs.harvard.edu/abs/1992ARA&A..30..575C} {30, 575}

\bibitem[\protect\citeauthoryear{{Condon}, {Cotton}  \& {Broderick}}{{Condon}
  et~al.}{2002}]{ccb02}
{Condon} J.~J.,  {Cotton} W.~D.,   {Broderick} J.~J.,  2002, \mn@doi [\aj]
  {10.1086/341650}, \href {http://adsabs.harvard.edu/abs/2002AJ....124..675C}
  {124, 675}

\bibitem[\protect\citeauthoryear{{Connor} et~al.,}{{Connor}
  et~al.}{2021}]{con21}
{Connor} T.,  et~al., 2021, \mn@doi [\apj] {10.3847/1538-4357/abe710}, \href
  {https://ui.adsabs.harvard.edu/abs/2021ApJ...911..120C} {911, 120}

\bibitem[\protect\citeauthoryear{{Fabian}, {Walker}, {Celotti}, {Ghisellini},
  {Mocz}, {Blundell}  \& {McMahon}}{{Fabian} et~al.}{2014}]{fg14}
{Fabian} A.~C.,  {Walker} S.~A.,  {Celotti} A.,  {Ghisellini} G.,  {Mocz} P.,
  {Blundell} K.~M.,   {McMahon} R.~G.,  2014, \mn@doi [\mnras]
  {10.1093/mnrasl/slu065}, \href
  {http://adsabs.harvard.edu/abs/2014MNRAS.442L..81F} {442, L81}

\bibitem[\protect\citeauthoryear{{Fan}, {Banados}  \& {Simcoe}}{{Fan}
  et~al.}{2022}]{fan23}
{Fan} X.,  {Banados} E.,   {Simcoe} R.~A.,  2022, \mn@doi []
  {10.48550/arXiv.2212.06907}, \href
  {https://ui.adsabs.harvard.edu/abs/2022arXiv221206907F} {p. arXiv:2212.06907}

\bibitem[\protect\citeauthoryear{{Ghisellini}, {Celotti}, {Tavecchio}, {Haardt}
   \& {Sbarrato}}{{Ghisellini} et~al.}{2014}]{gh14}
{Ghisellini} G.,  {Celotti} A.,  {Tavecchio} F.,  {Haardt} F.,   {Sbarrato} T.,
   2014, \mn@doi [\mnras] {10.1093/mnras/stt2394}, \href
  {http://adsabs.harvard.edu/abs/2014MNRAS.438.2694G} {438, 2694}

\bibitem[\protect\citeauthoryear{{Gloudemans} et~al.,}{{Gloudemans}
  et~al.}{2021}]{glou21}
{Gloudemans} A.~J.,  et~al., 2021, \mn@doi [\aap]
  {10.1051/0004-6361/202141722}, \href
  {https://ui.adsabs.harvard.edu/abs/2021A&A...656A.137G} {656, A137}

\bibitem[\protect\citeauthoryear{{Gloudemans} et~al.,}{{Gloudemans}
  et~al.}{2022}]{glou22}
{Gloudemans} A.~J.,  et~al., 2022, \mn@doi [\aap]
  {10.1051/0004-6361/202244763}, \href
  {https://ui.adsabs.harvard.edu/abs/2022A&A...668A..27G} {668, A27}

\bibitem[\protect\citeauthoryear{{G{\"u}ltekin}, {Cackett}, {Miller}, {Di
  Matteo}, {Markoff}  \& {Richstone}}{{G{\"u}ltekin} et~al.}{2009}]{gul09}
{G{\"u}ltekin} K.,  {Cackett} E.~M.,  {Miller} J.~M.,  {Di Matteo} T.,
  {Markoff} S.,   {Richstone} D.~O.,  2009, \mn@doi [\apj]
  {10.1088/0004-637X/706/1/404}, \href
  {https://ui.adsabs.harvard.edu/abs/2009ApJ...706..404G} {706, 404}

\bibitem[\protect\citeauthoryear{{G{\"u}ltekin}, {Cackett}, {King}, {Miller}
  \& {Pinkney}}{{G{\"u}ltekin} et~al.}{2014}]{gul14}
{G{\"u}ltekin} K.,  {Cackett} E.~M.,  {King} A.~L.,  {Miller} J.~M.,
  {Pinkney} J.,  2014, \mn@doi [\apjl] {10.1088/2041-8205/788/2/L22}, \href
  {https://ui.adsabs.harvard.edu/abs/2014ApJ...788L..22G} {788, L22}

\bibitem[\protect\citeauthoryear{{G{\"u}ltekin}, {King}, {Cackett}, {Nyland},
  {Miller}, {Di Matteo}, {Markoff}  \& {Rupen}}{{G{\"u}ltekin}
  et~al.}{2019}]{gul19}
{G{\"u}ltekin} K.,  {King} A.~L.,  {Cackett} E.~M.,  {Nyland} K.,  {Miller}
  J.~M.,  {Di Matteo} T.,  {Markoff} S.,   {Rupen} M.~P.,  2019, \mn@doi [\apj]
  {10.3847/1538-4357/aaf6b9}, \href
  {https://ui.adsabs.harvard.edu/abs/2019ApJ...871...80G} {871, 80}

\bibitem[\protect\citeauthoryear{{Herrington}, {Whalen}  \&
  {Woods}}{{Herrington} et~al.}{2023}]{herr23a}
{Herrington} N.~P.,  {Whalen} D.~J.,   {Woods} T.~E.,  2023, \mn@doi [\mnras]
  {10.1093/mnras/stad572}, \href
  {https://ui.adsabs.harvard.edu/abs/2023MNRAS.521..463H} {521, 463}

\bibitem[\protect\citeauthoryear{{Hosokawa}, {Yorke}, {Inayoshi}, {Omukai}  \&
  {Yoshida}}{{Hosokawa} et~al.}{2013}]{hos13}
{Hosokawa} T.,  {Yorke} H.~W.,  {Inayoshi} K.,  {Omukai} K.,   {Yoshida} N.,
  2013, \mn@doi [\apj] {10.1088/0004-637X/778/2/178}, \href
  {http://adsabs.harvard.edu/abs/2013ApJ...778..178H} {778, 178}

\bibitem[\protect\citeauthoryear{{Joggerst}, {Almgren}, {Bell}, {Heger},
  {Whalen}  \& {Woosley}}{{Joggerst} et~al.}{2010}]{jet09b}
{Joggerst} C.~C.,  {Almgren} A.,  {Bell} J.,  {Heger} A.,  {Whalen} D.,
  {Woosley} S.~E.,  2010, \mn@doi [\apj] {10.1088/0004-637X/709/1/11}, \href
  {http://adsabs.harvard.edu/abs/2010ApJ...709...11J} {709, 11}

\bibitem[\protect\citeauthoryear{{Kennicutt}}{{Kennicutt}}{1998}]{ken98}
{Kennicutt} Robert~C. J.,  1998, \mn@doi [\araa]
  {10.1146/annurev.astro.36.1.189}, \href
  {https://ui.adsabs.harvard.edu/abs/1998ARA&A..36..189K} {36, 189}

\bibitem[\protect\citeauthoryear{{K{\"o}rding}, {Falcke}  \&
  {Corbel}}{{K{\"o}rding} et~al.}{2006}]{kord06}
{K{\"o}rding} E.,  {Falcke} H.,   {Corbel} S.,  2006, \mn@doi [\aap]
  {10.1051/0004-6361:20054144}, \href
  {https://ui.adsabs.harvard.edu/abs/2006A&A...456..439K} {456, 439}

\bibitem[\protect\citeauthoryear{{Latif}, {Whalen}, {Khochfar}, {Herrington}
  \& {Woods}}{{Latif} et~al.}{2022a}]{latif22b}
{Latif} M.~A.,  {Whalen} D.~J.,  {Khochfar} S.,  {Herrington} N.~P.,   {Woods}
  T.~E.,  2022a, \mn@doi [\nat] {10.1038/s41586-022-04813-y}, \href
  {https://ui.adsabs.harvard.edu/abs/2022Natur.607...48L} {607, 48}

\bibitem[\protect\citeauthoryear{{Latif}, {Whalen}  \& {Khochfar}}{{Latif}
  et~al.}{2022b}]{latif22a}
{Latif} M.~A.,  {Whalen} D.,   {Khochfar} S.,  2022b, \mn@doi [\apj]
  {10.3847/1538-4357/ac3916}, \href
  {https://ui.adsabs.harvard.edu/abs/2022ApJ...925...28L} {925, 28}

\bibitem[\protect\citeauthoryear{{Lupi}, {Haiman}  \& {Volonteri}}{{Lupi}
  et~al.}{2021}]{lup21}
{Lupi} A.,  {Haiman} Z.,   {Volonteri} M.,  2021, \mn@doi [\mnras]
  {10.1093/mnras/stab692}, \href
  {https://ui.adsabs.harvard.edu/abs/2021MNRAS.503.5046L} {503, 5046}

\bibitem[\protect\citeauthoryear{{Marconi}, {Risaliti}, {Gilli}, {Hunt},
  {Maiolino}  \& {Salvati}}{{Marconi} et~al.}{2004}]{marc04}
{Marconi} A.,  {Risaliti} G.,  {Gilli} R.,  {Hunt} L.~K.,  {Maiolino} R.,
  {Salvati} M.,  2004, \mn@doi [\mnras] {10.1111/j.1365-2966.2004.07765.x},
  \href {http://adsabs.harvard.edu/abs/2004MNRAS.351..169M} {351, 169}

\bibitem[\protect\citeauthoryear{{Mayer}, {Kazantzidis}, {Escala}  \&
  {Callegari}}{{Mayer} et~al.}{2010}]{may10}
{Mayer} L.,  {Kazantzidis} S.,  {Escala} A.,   {Callegari} S.,  2010, \mn@doi
  [\nat] {10.1038/nature09294}, \href
  {http://adsabs.harvard.edu/abs/2010Natur.466.1082M} {466, 1082}

\bibitem[\protect\citeauthoryear{{McGreer}, {Becker}, {Helfand}  \&
  {White}}{{McGreer} et~al.}{2006}]{mcg06}
{McGreer} I.~D.,  {Becker} R.~H.,  {Helfand} D.~J.,   {White} R.~L.,  2006,
  \mn@doi [\apj] {10.1086/507767}, \href
  {https://ui.adsabs.harvard.edu/abs/2006ApJ...652..157M} {652, 157}

\bibitem[\protect\citeauthoryear{{Meiksin} \& {Whalen}}{{Meiksin} \&
  {Whalen}}{2013}]{mw12}
{Meiksin} A.,  {Whalen} D.~J.,  2013, \mn@doi [\mnras] {10.1093/mnras/stt089},
  \href {http://adsabs.harvard.edu/abs/2013MNRAS.430.2854M} {430, 2854}

\bibitem[\protect\citeauthoryear{{Merloni}, {Heinz}  \& {di Matteo}}{{Merloni}
  et~al.}{2003}]{merl03}
{Merloni} A.,  {Heinz} S.,   {di Matteo} T.,  2003, \mn@doi [\mnras]
  {10.1046/j.1365-2966.2003.07017.x}, \href
  {http://adsabs.harvard.edu/abs/2003MNRAS.345.1057M} {345, 1057}

\bibitem[\protect\citeauthoryear{{Mezcua}, {Hlavacek-Larrondo}, {Lucey},
  {Hogan}, {Edge}  \& {McNamara}}{{Mezcua} et~al.}{2018}]{mez18}
{Mezcua} M.,  {Hlavacek-Larrondo} J.,  {Lucey} J.~R.,  {Hogan} M.~T.,  {Edge}
  A.~C.,   {McNamara} B.~R.,  2018, \mn@doi [\mnras] {10.1093/mnras/stx2812},
  \href {https://ui.adsabs.harvard.edu/abs/2018MNRAS.474.1342M} {474, 1342}

\bibitem[\protect\citeauthoryear{{Momjian}, {Carilli}, {Walter}  \&
  {Venemans}}{{Momjian} et~al.}{2014}]{mom14}
{Momjian} E.,  {Carilli} C.~L.,  {Walter} F.,   {Venemans} B.,  2014, \mn@doi
  [\aj] {10.1088/0004-6256/147/1/6}, \href
  {https://ui.adsabs.harvard.edu/abs/2014AJ....147....6M} {147, 6}

\bibitem[\protect\citeauthoryear{{Momjian}, {Carilli}, {Ba{\~n}ados}, {Walter}
  \& {Venemans}}{{Momjian} et~al.}{2018}]{mom18}
{Momjian} E.,  {Carilli} C.~L.,  {Ba{\~n}ados} E.,  {Walter} F.,   {Venemans}
  B.~P.,  2018, \mn@doi [\apj] {10.3847/1538-4357/aac76f}, \href
  {https://ui.adsabs.harvard.edu/abs/2018ApJ...861...86M} {861, 86}

\bibitem[\protect\citeauthoryear{{Mortlock} et~al.,}{{Mortlock}
  et~al.}{2011}]{mort11}
{Mortlock} D.~J.,  et~al., 2011, \mn@doi [\nat] {10.1038/nature10159}, \href
  {http://adsabs.harvard.edu/abs/2011Natur.474..616M} {474, 616}

\bibitem[\protect\citeauthoryear{{Nanni}, {Hennawi}, {Wang}, {Yang},
  {Schindler}  \& {Fan}}{{Nanni} et~al.}{2022}]{nan22}
{Nanni} R.,  {Hennawi} J.~F.,  {Wang} F.,  {Yang} J.,  {Schindler} J.-T.,
  {Fan} X.,  2022, \mn@doi [\mnras] {10.1093/mnras/stac1944}, \href
  {https://ui.adsabs.harvard.edu/abs/2022MNRAS.515.3224N} {515, 3224}

\bibitem[\protect\citeauthoryear{{Pacucci}, {Ferrara}, {Volonteri}  \&
  {Dubus}}{{Pacucci} et~al.}{2015}]{pac15}
{Pacucci} F.,  {Ferrara} A.,  {Volonteri} M.,   {Dubus} G.,  2015, \mn@doi
  [\mnras] {10.1093/mnras/stv2196}, \href
  {http://adsabs.harvard.edu/abs/2015MNRAS.454.3771P} {454, 3771}

\bibitem[\protect\citeauthoryear{{Planck Collaboration} et~al.,}{{Planck
  Collaboration} et~al.}{2016}]{planck2}
{Planck Collaboration} et~al., 2016, \mn@doi [\aap]
  {10.1051/0004-6361/201525830}, \href
  {http://adsabs.harvard.edu/abs/2016A%26A...594A..13P} {594, A13}

\bibitem[\protect\citeauthoryear{{Plotkin} \& {Reines}}{{Plotkin} \&
  {Reines}}{2018}]{pr18}
{Plotkin} R.~M.,  {Reines} A.~E.,  2018, \mn@doi []
  {10.48550/arXiv.1810.06814}, \href
  {https://ui.adsabs.harvard.edu/abs/2018arXiv181006814P} {}

\bibitem[\protect\citeauthoryear{{Plotkin}, {Markoff}, {Kelly}, {K{\"o}rding}
  \& {Anderson}}{{Plotkin} et~al.}{2012}]{plot12}
{Plotkin} R.~M.,  {Markoff} S.,  {Kelly} B.~C.,  {K{\"o}rding} E.,   {Anderson}
  S.~F.,  2012, \mn@doi [\mnras] {10.1111/j.1365-2966.2011.19689.x}, \href
  {https://ui.adsabs.harvard.edu/abs/2012MNRAS.419..267P} {419, 267}

\bibitem[\protect\citeauthoryear{{Prandoni} \& {Seymour}}{{Prandoni} \&
  {Seymour}}{2014}]{ska1}
{Prandoni} I.,  {Seymour} N.,  2014, \mn@doi [] {10.48550/arXiv.1412.6942},
  \href {https://ui.adsabs.harvard.edu/abs/2014arXiv1412.6942P} {}

\bibitem[\protect\citeauthoryear{{Smidt}, {Whalen}, {Johnson}, {Surace}  \&
  {Li}}{{Smidt} et~al.}{2018}]{smidt18}
{Smidt} J.,  {Whalen} D.~J.,  {Johnson} J.~L.,  {Surace} M.,   {Li} H.,  2018,
  \mn@doi [\apj] {10.3847/1538-4357/aad7b8}, \href
  {http://adsabs.harvard.edu/abs/2018ApJ...865..126S} {865, 126}

\bibitem[\protect\citeauthoryear{{Spingola}, {Dallacasa}, {Belladitta},
  {Caccianiga}, {Giroletti}, {Moretti}  \& {Orienti}}{{Spingola}
  et~al.}{2020}]{spng20}
{Spingola} C.,  {Dallacasa} D.,  {Belladitta} S.,  {Caccianiga} A.,
  {Giroletti} M.,  {Moretti} A.,   {Orienti} M.,  2020, \mn@doi [\aap]
  {10.1051/0004-6361/202039458}, \href
  {https://ui.adsabs.harvard.edu/abs/2020A&A...643L..12S} {643, L12}

\bibitem[\protect\citeauthoryear{{Tenneti}, {Di Matteo}, {Croft}, {Garcia}  \&
  {Feng}}{{Tenneti} et~al.}{2018}]{ten18}
{Tenneti} A.,  {Di Matteo} T.,  {Croft} R.,  {Garcia} T.,   {Feng} Y.,  2018,
  \mn@doi [\mnras] {10.1093/mnras/stx2788}, \href
  {https://ui.adsabs.harvard.edu/abs/2018MNRAS.474..597T} {474, 597}

\bibitem[\protect\citeauthoryear{{Valentini}, {Gallerani}  \&
  {Ferrara}}{{Valentini} et~al.}{2021}]{vgf21}
{Valentini} M.,  {Gallerani} S.,   {Ferrara} A.,  2021, \mn@doi [\mnras]
  {10.1093/mnras/stab1992}, \href
  {https://ui.adsabs.harvard.edu/abs/2021MNRAS.507....1V} {507, 1}

\bibitem[\protect\citeauthoryear{{Vikaeus}, {Whalen}  \&
  {Zackrisson}}{{Vikaeus} et~al.}{2022}]{vik22a}
{Vikaeus} A.,  {Whalen} D.~J.,   {Zackrisson} E.,  2022, \mn@doi [\apjl]
  {10.3847/2041-8213/ac7802}, \href
  {https://ui.adsabs.harvard.edu/abs/2022ApJ...933L...8V} {933, L8}

\bibitem[\protect\citeauthoryear{{Wang} et~al.,}{{Wang} et~al.}{2021}]{wang21}
{Wang} F.,  et~al., 2021, \mn@doi [\apjl] {10.3847/2041-8213/abd8c6}, \href
  {https://ui.adsabs.harvard.edu/abs/2021ApJ...907L...1W} {907, L1}

\bibitem[\protect\citeauthoryear{{Whalen} \& {Fryer}}{{Whalen} \&
  {Fryer}}{2012}]{wf12}
{Whalen} D.~J.,  {Fryer} C.~L.,  2012, \mn@doi [\apjl]
  {10.1088/2041-8205/756/1/L19}, \href
  {http://adsabs.harvard.edu/abs/2012ApJ...756L..19W} {756, L19}

\bibitem[\protect\citeauthoryear{{Whalen}, {Abel}  \& {Norman}}{{Whalen}
  et~al.}{2004}]{wan04}
{Whalen} D.,  {Abel} T.,   {Norman} M.~L.,  2004, \mn@doi [\apj]
  {10.1086/421548}, \href {http://adsabs.harvard.edu/abs/2004ApJ...610...14W}
  {610, 14}

\bibitem[\protect\citeauthoryear{{Whalen}, {van Veelen}, {O'Shea}  \&
  {Norman}}{{Whalen} et~al.}{2008}]{wet08a}
{Whalen} D.,  {van Veelen} B.,  {O'Shea} B.~W.,   {Norman} M.~L.,  2008,
  \mn@doi [\apj] {10.1086/589643}, \href
  {http://adsabs.harvard.edu/abs/2008ApJ...682...49W} {682, 49}

\bibitem[\protect\citeauthoryear{{Whalen}, {Mezcua}, {Meiksin}, {Hartwig}  \&
  {Latif}}{{Whalen} et~al.}{2020a}]{wet20a}
{Whalen} D.~J.,  {Mezcua} M.,  {Meiksin} A.,  {Hartwig} T.,   {Latif} M.~A.,
  2020a, \mn@doi [\apjl] {10.3847/2041-8213/ab9a30}, \href
  {https://ui.adsabs.harvard.edu/abs/2020ApJ...896L..45W} {896, L45}

\bibitem[\protect\citeauthoryear{{Whalen}, {Surace}, {Bernhardt}, {Zackrisson},
  {Pacucci}, {Ziegler}  \& {Hirschmann}}{{Whalen} et~al.}{2020b}]{wet20b}
{Whalen} D.~J.,  {Surace} M.,  {Bernhardt} C.,  {Zackrisson} E.,  {Pacucci} F.,
   {Ziegler} B.,   {Hirschmann} M.,  2020b, \mn@doi [\apjl]
  {10.3847/2041-8213/ab9d29}, \href
  {https://ui.adsabs.harvard.edu/abs/2020ApJ...897L..16W} {897, L16}

\bibitem[\protect\citeauthoryear{{Whalen}, {Mezcua}, {Patrick}, {Meiksin}  \&
  {Latif}}{{Whalen} et~al.}{2021}]{wet21a}
{Whalen} D.~J.,  {Mezcua} M.,  {Patrick} S.~J.,  {Meiksin} A.,   {Latif} M.~A.,
   2021, \mn@doi [\apjl] {10.3847/2041-8213/ac35e6}, \href
  {https://ui.adsabs.harvard.edu/abs/2021ApJ...922L..39W} {922, L39}

\bibitem[\protect\citeauthoryear{{Willott} et~al.,}{{Willott}
  et~al.}{2010}]{wil10}
{Willott} C.~J.,  et~al., 2010, \mn@doi [\aj] {10.1088/0004-6256/139/3/906},
  \href {https://ui.adsabs.harvard.edu/abs/2010AJ....139..906W} {139, 906}

\bibitem[\protect\citeauthoryear{{Woods}, {Patrick}, {Elford}, {Whalen}  \&
  {Heger}}{{Woods} et~al.}{2021}]{tyr21a}
{Woods} T.~E.,  {Patrick} S.,  {Elford} J.~S.,  {Whalen} D.~J.,   {Heger} A.,
  2021, \mn@doi [\apj] {10.3847/1538-4357/abfaf9}, \href
  {https://ui.adsabs.harvard.edu/abs/2021ApJ...915..110W} {915, 110}

\bibitem[\protect\citeauthoryear{{Yang} et~al.,}{{Yang} et~al.}{2020}]{yang20}
{Yang} J.,  et~al., 2020, \mn@doi [\apjl] {10.3847/2041-8213/ab9c26}, \href
  {https://ui.adsabs.harvard.edu/abs/2020ApJ...897L..14Y} {897, L14}

\bibitem[\protect\citeauthoryear{{Zhang}, {An}, {Wang}, {Frey}, {Gurvits},
  {Gab{\'a}nyi}, {Perger}  \& {Paragi}}{{Zhang} et~al.}{2022}]{zng22}
{Zhang} Y.,  {An} T.,  {Wang} A.,  {Frey} S.,  {Gurvits} L.~I.,  {Gab{\'a}nyi}
  K.~{\'E}.,  {Perger} K.,   {Paragi} Z.,  2022, \mn@doi [\aap]
  {10.1051/0004-6361/202243785}, \href
  {https://ui.adsabs.harvard.edu/abs/2022A&A...662L...2Z} {662, L2}

\makeatother
\end{thebibliography}






\bsp	
\label{lastpage}

\end{document}